\documentstyle[12pt]{article}
\def\a{\alpha}
\def\b{\beta}
\def\rar{\rightarrow}

\def\bpsi{\bar{\psi}}

\def\dg{{\dagger}}

\def\le{\left(}
\def\ri{\right)}
\def\t{\theta}
\def\bt{{\bar{\theta}}}

\def\ve{\varepsilon}
\def\bve{{\bar{\varepsilon}}}
\def\wx{\tilde{x}} 
\def\wt{\tilde{\theta}}
\def\wbt{{\bar{\tilde{\theta}}}}
\def\Qb{{\bar{Q}}}
\def\pd{\partial}

\def\ol{\overleftarrow}
\def\Db{{\bar{D}}}
\def\g{{\gamma}}

\def\Dc{{\cal D}}
\def\f12{\frac{1}{2}}

\def\dis{\displaystyle}
\def\del{\delta}
\def\mub{{\bar{\mu}}}
\begin{document}
\begin{titlepage}
\flushright{SISSA/30/99/EP}

\vspace{2cm}
\begin{center}
{\Large \bf  On the relation between  Green's functions \\ 
\vspace{3mm} 
of the SUSY theory with and without soft terms}

\vglue 10mm
 Igor Kondrashuk\footnote{E-mail: ikond@sissa.it, 
on leave of absence from LNP, JINR, Dubna, Russia}

\vglue 5mm
{\it SISSA -- ISAS  and INFN, Sezione di Trieste, \\
 Via Beirut 2-4, I-34013, Trieste, Italy}

\end{center}

\vglue 20mm
\begin{abstract}
We study possible relations between the full Green's functions of 
softly broken supersymmetric theories and the full Green's functions 
of rigid supersymmetric theories on the example of the supersymmetric 
quantum mechanics and find that algebraic relations can exist and can be 
written in a simple form. These algebraic relations between the Green's 
functions  have been derived by transforming the path integral of the rigid theory. 
In this approach soft terms appear as the result of general 
changes of coordinates in the superspace. 
\end{abstract}


\end{titlepage}

The renormalization of softly broken supersymmetric Yang-Mills theories
in four space-time dimensions into the framework of the superfield 
technique and spurions method was studied from different points of view 
in Refs. (\cite{Yam}-\cite{Giudice-2}). The soft terms were introduced 
as the finite series in terms of Grassmanian coordinates of the 
superspace. It results in dependence of renormalization 
constants of the theory on the Grassmanian coordinates. As it was 
understood, there are direct relations between the renormalization 
constants of the rigid theory -- theory without soft terms -- and  
the renormalization constants of its softly broken counterpart. 
The first investigation in this direction was performed 
by Yamada \cite{Yam} and it was developed in \cite{Jones,AKK},
where simple rules to calculate these constants starting from 
the well-known results for the rigid theory have been derived. These rules are 
realized in terms of differential operators acting in the coupling 
space of the rigid theory. In the Ref.\cite{AKK} it was found that 
this approach is in good agreement with nonperturbative 
results \cite{Shifman}. The reason why these rules exist is that 
the soft breaking terms can be considered as  background 
$x$-independent superfields \cite{AKK,Giudice-2} in which our theory 
is embedded. These external superfields must be substituted  
into infinite parts of the Green's functions instead of the 
couplings of the rigid theory. The algebraic relations
between the renormalizations constants are consequences
of these substitutions \cite{AKK}.    

However, it would be nice to have relations for the complete 
Green's functions of the theory that includes soft terms and 
for its counterpart without them. Here we find these relations 
for the case of the supersymmetric quantum mechanics as a simplest 
example where the method proposed below can work. The supersymmetric 
quantum mechanics was developed in (\cite{SQM1}-\cite{SQM3})
and its action is \footnote{Here the superscript $R$ stands 
for ``rigid''}

\begin{eqnarray}
S^{R}  =   \int dx d\t d\bt \le D\Phi\Db\Phi  + V(\Phi) \ri, \label{RidS}
\end{eqnarray}
where $V$ is an arbitrary function of the real superfield  $\Phi,$

\begin{eqnarray*}
\Phi\le x,\t,\bt \ri =  \phi(x) + i\t\psi(x) + i\bt\bpsi(x) + \t\bt D(x), ~~~ 
\Phi^\dg = \Phi.
\end{eqnarray*}

The Lagrangian (\ref{RidS}) is the only thing of the supersymmetric 
quantum mechanics we need. In the rest of this letter we work  with 
the path integral produced by this classical action. The Green's functions of 
it are referred in what follows as the rigid Green's functions: 

\begin{eqnarray*}
& \dis{G_n^{R}\le x_1,x_2,\dots, x_n,\t_1,\dots,\t_n,\bt_1,\dots,\bt_n\ri} = \\
& \dis{\frac{\del}{\del J(x_1,\t_1,\bt_n)}
 \frac{\del}{\del J(x_2,\t_1,\bt_1)}\dots\frac{\del}{\del J(x_n,\t_n,\bt_n)}
Z^R[J]}|_{J=0}, \\
&  \dis{Z^R[J] = \int{\Dc\Phi} \exp\le S^{R} + \int dx d\t d\bt J\Phi \ri}.
\end{eqnarray*}
Some comments about this expression are necessary. The Green's functions are 
local in the Grassmanian $\t$-coordinates \cite{West}. This result can be seen
in the perturbation theory after performing the Grassmanian integration. However,
we would like to keep all the Grassmanian coordinates in the above expression 
since all the following speculations are based on the path integral only. 
The external source $J$ is also real superfield, $J^\dg = J.$ The covariant 
derivatives in the action (\ref{RidS}) are defined as 
\begin{eqnarray*} 
D = \frac{\pd}{\pd\t} - i\bt\frac{\pd}{\pd x}, ~~~ 
\Db = \ol{\frac{\pd}{\pd\bt}} + i\ol{\t}\frac{\pd}{\pd x}.
\end{eqnarray*}
Polynomials in Grassmanian coordinates which are background 
$x$-independent superfields can stand in front of the terms 
of the action (\ref{RidS}), 
\begin{eqnarray}
& \dis{S^{S}  =   \int dx d\t d\bt \le  P_1(\t,\bt)D\Phi\Db\Phi  +  
P_2(\t,\bt)V(\Phi) \ri},  \nonumber\\
& P_i (\t,\bt) = \a_i + i\b_i\t + i\bar{\b}_i\bt + \g_i\t\bt \label{POLY}, 
~~~ P_i^\dg = P_i.
\end{eqnarray}
Here $\a_i$, $\b_i$, and $\g_i$ are arbitrary numbers, $\b_i$
is Grassmanian. In this case we say that there is a soft 
supersymmetry breaking since if we do not transform charges 
in a proper way we have non-invariance of the component action under the 
supersymmetry transformations caused by the change of the coordinates
in the superspace 
\begin{eqnarray}
& \dis{x \rar \exp\le\ve Q + \Qb\bve\ri x,~~~ 
\t \rar \t + \ve,~~~ \bt \rar \bt + \bve} \nonumber\\
& \dis{Q = \frac{\pd}{\pd\t} + i\bt\frac{\pd}{\pd x},~~~ 
\Qb = \ol{\frac{\pd}{\pd\bt}} - i\ol{\t}\frac{\pd}{\pd x}}. \label{SUSY}   
\end{eqnarray}
Our purpose is to show a way in which the rigid Green's functions
and the soft Green's functions, 
\begin{eqnarray*}
& \dis{G_n^{S}\le x_1,x_2,\dots, x_n,\t_1,\dots,\t_n,\bt_1,\dots,\bt_n\ri} = \\
& \dis{\frac{\del}{\del J(x_1,\t_1,\bt_1)}
 \frac{\del}{\del J(x_2,\t_2,\bt_2)}\dots\frac{\del}{\del J(x_n,\t_n,\bt_n)}
Z^S[J]}|_{J=0}, \\
&  \dis{Z^S[J] = \int{\Dc\Phi} \exp\le S^{S} + \int dx d\t d\bt J\Phi \ri},
\end{eqnarray*}
can be related. First of all, let us suppose that a general change of 
the superspace coordinates is made, 
\begin{eqnarray}
x = x(\wx, \wt, \wbt), ~~~ \t = \t(\wx, \wt, \wbt),~~~ 
\bt = \bt (\wx, \wt, \wbt),
\label{GT}
\end{eqnarray}
and we look for changes which satisfy the conditions 
\begin{eqnarray}
& \Db\le \wx - i\wt\wbt \ri = 0, ~~~ D\le \wx + i\wt\wbt\ri = 0, \nonumber\\ 
& \Db\wt =0, ~~~ D\wbt = 0, ~~~ \wx^\dg = \wx, \label{Cs}
\end{eqnarray}
where we suppose that we know the reversed law of the transformation
(\ref{GT})
\begin{eqnarray*}
\wx = \wx(x,\t,\bt), ~~~ \wt = \wt(x,\t,\bt),~~~ \wbt = \wbt (x,\t,\bt).
\end{eqnarray*}
The conditions (\ref{Cs}) mean that the covariant derivative 
transforms as 
\begin{eqnarray} 
D =  \frac{\pd}{\pd\t} - i\bt\frac{\pd}{\pd x} \rar \le D\wt \ri
\le\frac{\pd}{\pd\wt} - i\wbt\frac{\pd}{\pd \wx} \ri, \label{B}
\end{eqnarray}
and $\Db$ transforms in the Hermitian conjugated way. The most general change of 
superspace coordinates that satisfy the condition (\ref{Cs}) is 
\begin{eqnarray} 
& \wx = g(x) + i\mub\t f(x)\sqrt{\pd g(x)} + i\mu\bt f(x)\sqrt{\pd g(x)} 
- \mu\mub\pd f^2(x)\t\bt \nonumber\\
& \wt = \t\sqrt{\pd g(x - i\t\bt)} + \mu f(x - i\t\bt) \label{L}\\
& \wbt = \bt\sqrt{\pd g(x + i\t\bt)} + \mub f(x + i\t\bt)\nonumber ,
\end{eqnarray}
where the functions $f$ and $g$ are arbitrary and $\mu$ is a Grassmanian
dimensionful constant. The Berezinian of this change of coordinates 
is a long expression and we do not write it here for the brevity. 
Nevertheless, it is possible to see already in the law (\ref{L})
that the $(\t = \bt = 0)$-component of the Berezinian is equal to 1
for any functions $f$ and $g.$ At the same time the factors 
$D\wt$ and $\Db\wbt$ arising in (\ref{B}) under the change of coordinates 
in front of the covariant derivatives in the kynetic term of 
the rigid action $S^R$ has the $\sqrt{\pd g}$ as its   
$(\t = \bt = 0)$-component. Therefore, in order to have the 
independent on the space coordinate factor after the change is made
we should  take $g(x) = x$ and keep $f(x)$ arbitrary for a moment.
Then the form of the above transformation is 
\begin{eqnarray} 
& \wx = x + i\mub\t f(x) + i\mu\bt f(x) 
- \mu\mub\pd f^2 (x)\t\bt \nonumber\\
& \wt = \t + \mu f(x - i\t\bt) \label{TR2}\\
& \wbt = \bt + \mub f(x + i\t\bt)\nonumber. 
\end{eqnarray}

The change of coordinates  reversed to (\ref{TR2}) is 
\begin{eqnarray} 
& x = \wx - i\mub\wt f(\wx) - i\mu\wbt f(\wx)
- \mu\mub\pd f^2 (\wx)\wt\wbt \nonumber\\
& \t = \wt\le 1 + i\mu\mub\pd f^2 (\wx)\ri -  
\mu f(\wx - i\wt\wbt)  \label{TR3}\\
& \bt = \wbt \le 1 - i\mu\mub\pd f^2 (\wx)\ri 
- \mub f(\wx + i\wt\wbt) \nonumber. 
\end{eqnarray}
By the explicit calculation we see that the Berezinian of this change 
is equal to 1 for any function $f.$ To have a common factor arising in 
(\ref{B}) as $x$-independent background superfield, we should choose 
$f(x) = x.$ In this case, the transformation (\ref{TR2}) is 
\begin{eqnarray} 
& \wx = x\le 1  + i\mub\t  + i\mu\bt  
- 2 \mu\mub \t\bt \ri = x\le 1 - i\mub\t - i\mu\bt\ri^{-1}\nonumber\\
& \wt = \t + \mu (x - i\t\bt) \label{TR4}\\
& \wbt = \bt + \mub (x + i\t\bt), \nonumber 
\end{eqnarray}
and the reversed transformation is 
\begin{eqnarray} 
& x = \wx\le 1 + i\mub\wt + i\mu\wbt\ri^{-1} \nonumber\\
& \t = \le\wt - \mu (\wx - i\wt\wbt)\ri \exp(-2i\mub\wt) \label{TR5}\\
& \bt = \le\wbt - \mub (\wx + i\wt\wbt)\ri \exp(-2i\mu\wbt) \nonumber. 
\end{eqnarray}
The factors $D\wt$ and $\Dc\wbt$ that come from (\ref{B}) are 
\begin{eqnarray}
D\wt = \exp\le 2i\mu\bt \ri = \exp \le 2i\mu[\wbt 
  - \mub(\wx + i\wt\wbt)]\ri, \nonumber\\  
\Db\wbt = \exp\le 2i\mub\t \ri = \exp \le 2i\mub[\wt 
  - \mu(\wx - i\wt\wbt)]\ri. \label{FA}
\end{eqnarray}
Therefore, the common factor that is due to the transformation of 
the derivatives (\ref{B}) is 
\begin{eqnarray}
D\wt \Db\wbt = \exp \le 2i\mu\wbt +  2i\mub\wt + 4\mu\mub\wt\wbt \ri = 
1 + 2i\mu\wbt + 2i\mub\wt \label{CF}
\end{eqnarray}
 
We can make the change of coordinates (\ref{TR5}) in the rigid action 
$S^R$ of the path integral $Z^R$:
\begin{eqnarray}
& \dis{Z^R[J] = \int{\Dc\Phi} \exp\le \int dx d\t d\bt 
[ D\Phi\Db\Phi  + V(\Phi) + J\Phi]\ri = } \label{REL}\\
& = \dis{\int{\Dc\tilde{\Phi}}} \exp\le \dis{\int dx d\t d\bt 
[(1 + 2i\mu\bt + 2i\mub\t)D\tilde{\Phi}\Db\tilde{\Phi}}  
+ \dis{V(\tilde{\Phi})
+ \tilde{J}\tilde{\Phi}]}\ri = Z^S[\tilde{J}],\nonumber
\end{eqnarray}
where the definitions are used: 
\begin{eqnarray*}
& \dis{\tilde{\Phi}(x,\t,\bt) = \Phi\le \frac{x}{1 + i\mub\t +  i\mu\bt},~
\le\t - \mu (x - i\t\bt)\ri e^{-2i\mub\t},~ 
\le\bt - \mub (x + i\t\bt)\ri e^{-2i\mu\bt}\ri,} \\
& \dis{\tilde{J}(x,\t,\bt) \equiv J\le\frac{x}{1 + i\mub\t +  
i\mu\bt},~
\le\t - \mu (x - i\t\bt)\ri e^{-2i\mub\t},~ 
\le\bt - \mub (x + i\t\bt)\ri e^{-2i\mu\bt}\ri.}
\end{eqnarray*}

As one can see the second part of (\ref{REL}) is the path integral of 
the theory with soft supersymmetry breaking terms that correspond 
according to  (\ref{POLY}) to 
\begin{eqnarray}
& P_1 (\t,\bt) = 1 + 2i\mu\bt + 2i\mub\t,~~~
P_2 (\t,\bt) = 1 \label{POLY2}
\end{eqnarray}
We can expand this soft path integral $Z^S$ in terms of the external 
source $\tilde{J}$ and the coefficient functions of this expansion 
are in fact the Green's functions of the soft theory 
\begin{eqnarray}
& \dis{Z^S[\tilde{J}] = \sum_n\int dx_1dx_2\dots dx_n d\t_1\dots d\t_n 
d\bt_1\dots d\bt_n\frac{1}{n!} }
\dis{G_n^S\le x_1, x_2,\dots, x_n,\t_1,\t_2,\dots, \right.} \nonumber\\
& \left. \dots,\t_n,\bt_1,\bt_2,\dots,\bt_n\ri
\tilde{J}(x_1,\t_1,\bt_1)\tilde{J}(x_2,\t_2,\bt_2)\dots 
\tilde{J}(x_n,\t_n,\bt_n) \label{Inter}
\end{eqnarray}
Now we can make the change of coordinates back and restore the 
original sources $J$. It means that by changing the coordinates in the 
superspace as in (\ref{TR4}) 
for each of $x_i$ we have  instead of (\ref{Inter}) the expression 
\begin{eqnarray}
& \dis{Z^S[\tilde{J}[J]] = \sum_n\int dx_1dx_2\dots dx_n d\t_1\dots d\t_n 
d\bt_1\dots d\bt_n\frac{1}{n!}} *\nonumber \\ 
& \dis{G_n^S\le \frac{x_1}{1 - i\mub\t_1 - i\mu\bt_1},\dots, 
\frac{x_n}{1 - i\mub\t_n  - i\mu\bt_n},\t_1 + \mu (x_1 - i\t_1\bt_1),\dots,  
\right.} \nonumber\\
& \dots, \left. \t_n + \mu (x_n - i\t_n\bt_n),\bt_1 + \mub (x_1 + i\t_1\bt_1),
\dots,\bt_n + \mub (x_n + i\t_n\bt_n) \ri*\nonumber\\
& *J(x_1,\t_1,\bt_1)J(x_2,\t_2,\bt_2)\dots J(x_n,\t_n,\bt_n). \label{Final}
\end{eqnarray}

As one can see from (\ref{REL}), the expansion (\ref{Final}) is at the 
same time the expansion of the $Z^R[J]$ in terms of the external source 
of the rigid theory $J.$ Therefore, the following equality takes 
place 
\begin{eqnarray}
& G_n^R\le x_1, x_2,\dots, x_n,\t_1,\t_2,\dots,\t_n,\bt_1,\bt_2,\dots,\bt_n\ri
= \nonumber\\
& \dis{G_n^S\le \frac{x_1}{1 - i\mub\t_1 - i\mu\bt_1},\dots, 
\frac{x_n}{1 - i\mub\t_n  - i\mu\bt_n},\t_1 + \mu (x_1 - i\t_1\bt_1),\dots,  
\right.} \nonumber\\
& \dots, \left. \t_n + \mu (x_n - i\t_n\bt_n),\bt_1 + \mub (x_1 + i\t_1\bt_1),
\dots,\bt_n + \mub (x_n + i\t_n\bt_n) \ri. \label{REL2}
\end{eqnarray}

Having used the substitutions (\ref{TR5}) we transform the equation 
(\ref{REL2}) to the form 

\begin{eqnarray}
& G_n^S\le x_1, x_2,\dots, x_n,\t_1,\t_2,\dots,\t_n,\bt_1,\bt_2,\dots,\bt_n\ri
= \nonumber\\
& \dis{G_n^R\le \frac{x_1}{1 + i\mub\t_1 + i\mu\bt_1},\dots, 
\frac{x_n}{1 + i\mub\t_n  + i\mu\bt_n},
\le\t_1 - \mu (x_1 - i\t_1\bt_1)\ri e^{-2i\mub\t_1},\dots,  
\right.} \nonumber\\
& \dots, \left. \dis{\le\t_n - \mu (x_n - i\t_n\bt_n)\ri e^{-2i\mub\t_n},
\le\bt_1 - \mub (x_1 + i\t_1\bt_1)\ri e^{-2i\mu\bt_1},
\dots,} \right.\nonumber\\
& \dots, \left. \dis{\le \bt_n - \mub (x_n + i\t_n\bt_n)\ri e^{-2i\mu\bt_n}} \ri. 
\label{REL3}
\end{eqnarray}

Thus, the final result is that the theory with soft supersymmetry breaking
terms (\ref{POLY}) in the form (\ref{POLY2}) is equivalent to the 
rigid theory (\ref{RidS}) in the sense of the relation (\ref{REL3})
between their Green's functions. It looks like something surprising that 
we could relate soft and rigid theories, but the explanation is 
if we amount couplings to background $x$-independent superfields 
we have their transformations at the level of the component action
which are the reflection of the change of coordinates in the superspace
(\ref{SUSY}). Hence, if we treat couplings of theory at the component
level as components of an external background multiplet,  
we have no supersymmetry breaking since we include rigid couplings
and soft couplings in the supersymmetry transformation (\ref{SUSY}). 
Nevertheless, we can create soft terms under general changes  of the 
coordinates in the superspace. 

I thank  A. Masiero for stimulating discussions. The investigation is 
supported by INFN.

\end{document}